# EVIDENCE FOR SCALE FACTOR OSCILLATIONS OBSERVED IN THE LARGE SCALE STRUCTURE OF THE UNIVERSE

H. I. Ringermacher and L. R. Mead
Dept. of Physics and Astronomy, U. of Southern Mississippi, Hattiesburg, MS 39406, USA; ringerha@gmail.com, lawrence.mead@usm.edu

**ABSTRACT**

   We present two independent analyses as further evidence that galaxy clustering at scales of 500 Mpc and greater has a periodic time component induced by oscillations in the scale factor at a frequency of approximately 7 cycles over one Hubble time. The scale factor oscillations were discovered in previous work by analyzing the Hubble diagram for type 1a SNe data. In the present work we analyze galaxy number count data from SDSSIII-BOSS, DR9 using a simple oscillating expanding space model and also perform a Fourier analysis of the same SDSSIII data set . The number distribution of galaxies on these scales should be relatively smooth. However, a DR9 plot of galaxy number count per 0.01 redshift bin as a function of redshift shows significant bumps to redshift 0.5. Later releases show the same behavior. Our model fits essentially all bumps at 99.8% confidence once the oscillation is included. A Fourier analysis of the same number count vs. redshift data (processed only to convert redshift to equal time bins) clearly shows the dominant 7-cycle signal at 15/1 signal-to-noise ratio. The DR9 galaxy number count peaks near redshift 0.5 and then falls off due to target magnitude limitations. In our model we assume ideal observation to all redshifts.  The oscillation model displays a matching peak at redshift 0.5, then falls, but continues on to rise predicting a second peak at redshift 0.64. Confirmation of the second peak from future SDSSIV data to higher redshift would further support our observation of oscillations in the scale factor. The oscillations may derive from a scalar field model of dark matter as shown in our earlier work.

## I. INTRODUCTION

   In an earlier series of papers [1,2] we describe the observation of discrete oscillations in the scale factor as a function of cosmological time analyzed from an extensive set of type 1a supernovae (SNe) data [3] These temporal oscillations are not to be confused with Baryon Acoustic Oscillations (BAO) which left spatial imprints in the CMB generated at the fixed time of recombination. The period of our oscillations is very long - superseding the relatively short scale of reported periodicities in the red shift of galaxies [4,5] - well beyond the filament/void structure scale.  The fundamental frequency was measured to be 6.5 ± 0.6 cycles over one Hubble-time (13.8 Gyr) for which we coined the term 6.5 HHz (Hubble-Hertz) for convenience. Second and third harmonics of this fundamental at approximately 13 HHz and 20 HHz were also reported. The measured oscillations occurred between 3 and 7 billion years ago. Three different analyses, Gaussian filtering, Fourier transform and autocorrelation confirmed the fundamental and harmonics as sharp frequencies. The signal-to-noise ratio varied between 2 and 3 representing at least a 95% confidence level that the signal is real. A noise study, also presented in the paper, indicated a 5% chance that the signal could be generated from noise alone – a result consistent with the findings. Nevertheless, further supporting evidence is essential, either from independent new SNe data or other predicted consequences of such oscillations. In particular, in the discussion of [2] we predicted there should be an observable effect on clustering in the large scale structure (LSS) of the universe arising from these same frequencies. In this paper we report evidence for precisely such an effect at a greater than 3σ confidence level.

## 2. RESULTS

Figure 1 (dots) shows galaxy number distribution data, nDat(z), downloaded from the SDSSIII-BOSS, DR9 data release site (displayed as Figure 2 in [6]), describing galaxy number count per 0.01 redshift bin as a function of redshift. The count is a sum over the number of observed galaxies in 0.01 redshift bins measured out to redshift z = 0.8 (~7 Gyr lookback time) observed in various directional pie-wedges out from the Galaxy. Comments regarding DR9 [6] indicate this is a "near-uniform sampling" of number count over the total volume of the survey. Approximately 450,000 galaxies are counted. Figure 14 of DR10 Boss [8] is virtually identical but includes nearly twice as many galaxies. There are several points to note regarding the plot. The curve falls off past z = 0.5 because of target galaxy magnitude limitations. Using a simple physical model, described in Section 2.1, where galaxy number count is proportional to the volume of the expanding space as the cube of the coordinate, or more generally, luminosity distance, one would expect the curve to continue rising rapidly. One sees, however, a number of "bumps" in the curve. These are presumably due to clustering of galaxies at various redshifts, in particular between z = 0.2 – 0.4 and possibly at z = 0.5. Since galaxy number count is an averaging over statistically significant numbers of galaxies [9], and the data set covering ¼ of the sky represents a near-uniform sampling, the number count is expected to be isotropic and homogeneous over the stated scales across the sky. Yet significant bumps appear. This prompted the question – could it be that this isotropic clustering is induced by the scale factor oscillations? We use our analysis in [2] to create a simple model of galaxy number distribution. In addition to fitting our expansion model, which includes the fundamental 7 HHz scale factor oscillation, we perform a direct Fourier analysis of the raw galaxy number count vs. cosmological time data set to confirm the presence of the 7 HHz signal in Section 2.2.

### 2.1. Model Describing Large Scale Structure

#### 2.1.1 Definitions

The FRW flat space metric for the ΛCDM model independent of angle is given by

$$ds^2 = dt^2 - a(t)^2 dr^2 \qquad (1)$$

where $r$ is the coordinate distance or "comoving" distance. The scale factor, $a(t)$, is the radius of the universe normalized to 1 at the present time.

In [2] we showed that ΛCDM precisely fit our scalar field oscillation model as the mean. We then developed a simulated data model that used the scale factor from ΛCDM added to an oscillating component. The model scale factor is given by:

$$a(t) = a(t)_{\Lambda CDM} + \varepsilon \left(1.0\sin(\omega_1 t + \phi_1) + 0.41\sin(\omega_2 t + \phi_2) + 0.09\sin(\omega_3 t + \phi_3)\right)e^{-2.82t}. \quad (2)$$

The polynomial $a(t)_{\Lambda CDM} = -0.6782t^3 + 1.6032t^2 - 0.3345t + 0.4026$ is the mean of the scale factor data used in [2] and is a best fit to the standard model ΛCDM scale factor over the analyzed cosmological time range, $t = 0.4$ to $t = 1.0$. Frequency is defined here as $\omega = 2\pi f$, where $f$ is in units of Hubble-Hertz (HHz), coined in [2], and defined as 1 HHz =1 cycle over 1 Hubble-time. The frequency parameters are;

$$\omega_1 = 43.68, \; \omega_2 = 2\omega_1, \; \omega_3 = 3\omega_1. \quad (3)$$

The phase shifts used for best fit are $\phi_1 = -0.16, \phi_2 = 0.51, \phi_3 = 0.05$. These phases produce an oscillation starting at $t = 0$. The oscillation amplitude is consistent with that of [2], within the noise, and is $\varepsilon = 0.0302$.

Cosmological time is given by

$$t = 1 - \int_0^z \frac{dz'}{(1+z')E(z')}, \quad (4)$$

where, $E(z) = \sqrt{\Omega_m(1+z)^3 + \Omega_\Lambda}$.

$\Omega_m = 0.29$ is the present baryonic plus dark matter density parameter. $\Omega_\Lambda = 0.71$ is the dark energy density parameter. We require $a(1) = 1$ and $a'(1) = 1$ from [2].

Along a light-line the coordinate distance is given by

$$r(t) = \int_t^1 \frac{dt'}{a(t')}. \quad (5)$$

This form of the coordinate distance allows for the oscillating model $a(t)$.

We define two redshifts, the ΛCDM redshift, $z$, from

$$1 + z = \frac{1}{a(t)_{\Lambda CDM}} \quad (6)$$

and a redshift, $\tilde{z}$, based on our model, from

$$1 + \tilde{z} = \frac{1}{a(t)}. \quad (7)$$

The luminosity distance is given by

$$D_L(z) = r(t)/a(t). \quad (8)$$

We take our number count model from Ostriker [9]. As Ostriker points out, number count smoothes over the large scale structure. In this model the galaxy number count relies simply on a spatial volume expansion from the first bin at $z_1 = \Delta z = 0.01$. Number density is assumed constant [9,10]. Number count is given by

$$N(z) = \left(\frac{N(z_1)}{D_L(z_1)^3}\right) D_L(z)^3, \qquad (9)$$

We assume this number count as representative of the entire sky, though it was conducted over ¼ of the sky. This is reasonable in light of the isotropy of the CMB, the Cosmological Principle and computer simulations of the LSS at the largest scales.

We may write for the number per 0.01 redshift bin;

$$n(z) = const \frac{dN(z)}{d\tilde{z}}, \qquad (10)$$

In carrying out this calculation we take a numerical derivative of Eq. (9) directly with respect to $\tilde{z}$ for consistency with our model, but $n(z)$ is plotted with respect to the fiducial redshifts to compare with the downloaded data. $Const = 1748$ for a best fit.

### 2.2.2 Calculation procedure

For the initial model procedure we use the parameters, frequencies, phases and amplitudes, as defined in Section 2.1.1 only for the fundamental frequency, $\omega_1 = 43.68$. The other two amplitudes are set to zero. We begin by defining a set of 100 "fiducial" redshifts, z, corresponding to the 100 number-count bins, each 0.01, to $z = 1$, downloaded from DR9. Our DR9 data is available in [11]. A cosmological time for each $z$ is calculated from equation (4) and is introduced into equation (5) to calculate the coordinate distance, $r(t)$. This is a direct integration to the given time using the functional form of $a(t)$ from (2). We can now calculate the luminosity distance, (8), for each fiducial redshift, *z*. From luminosity distance we calculate the volumetric galaxy number count, (9). Since we have defined our model redshift, $\tilde{z}$, from (7), for each time *t*, we take the numerical derivative of $N(z)$ with respect to $\tilde{z}$ generating the initial fitting curve, $n(z)$, plotted in Figure 1 against fiducial redshift together with the SDSS data.

### 2.2.3 Time-shift and final fit

In Figure 1 the dashed curve describes the above model, with no time shifting but including only the fundamental at approximately 7 HHz. It has been scaled with $const = 1748$ so the "bump" amplitude roughly matches the data and thus provides a reference point for the fit. $n(z)$ in Figure 1 (dashed curve) looks tantalizingly similar to the SDSS data set (dots) were it not for the *z*-displacement. We therefore attempt to shift *z* to the left. In other words, our model process took place at too high a redshift, or too long ago in lookback time. We can create a time-shift by examining lookback time:

$$t_{lookback} = \int_0^{\tilde{z}} \frac{dz'}{(1+z')E(z')} \qquad (11)$$

We use ΛCDM lookback time because that is sufficient to 1$^{st}$ order in $\varepsilon$ for the shift evaluation. We can write this, for a small time shift $\Delta t$ from (11), as

$$\bar{z} = z + \Delta t (1+z)\sqrt{0.29(1+z)^3 + 0.71}. \qquad (12)$$

$\bar{z}$ is then inserted in place of z in the integral limit for the cosmological time calculation, Eq.(4). This generates a time shift, $\Delta t$, which is adjusted to match the curve to the data. The solid curve is thus time-shifted by $\Delta t = 0.1209$ Hubble-time. This produces a rough match. Adding the second and third harmonics at the indicated frequencies (3) generates the fit of Fig. 2. The smooth dotted curve in Figure 2 is the present model without any oscillation – the ΛCDM model limit.

The time shift necessary for a best match is a time delay between the appearance of the scale factor change and the formation of LSS. The scale factor governs the acceleration and deceleration of the universe. Acceleration discourages LSS formation while deceleration encourages LSS. The time delay is necessary because we do not use a full, computed, physical model including gravitation which would include distributed time effects during LSS formation.

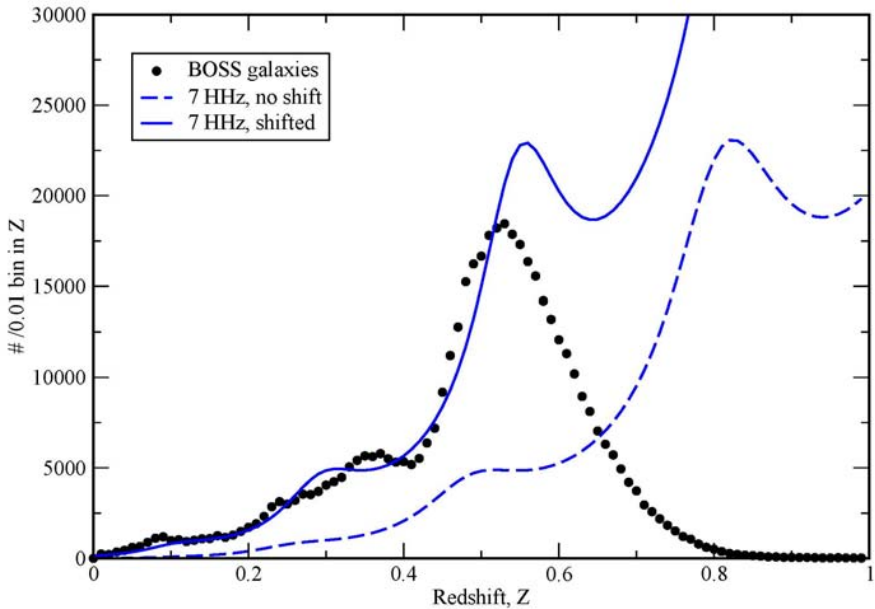

Fig.1 Galaxy number distribution data, nDat(z), (# galaxies/0.01 redshift bin), from SDSSIII-BOSS, DR9 (dots). Unshifted model curve with 7HHz fundamental only added (dashed). Time-shifted model curve, $n(z)$, (solid).

The $R^2$ goodness of fit is 0.998, equivalent to a 3.1 σ confidence level, to $z = 0.56$. Eq. (14) describes the goodness of fit analysis. $z_f$ is the last redshift for the fit, at $z = 0.56$, just past the data peak. *mean* is the mean of $nDat(z)$, (see Fig.1), between $z = 0$ and $z = 0.56$.

$$Rsquare(z_f) = 1 - \frac{\sum_{z=0}^{z_f} (n(z) - nDat(z))^2}{\sum_{z=0}^{z_f} (nDat(z) - mean)^2} \qquad (14)$$

The residuals between model and data appear random, supporting $R^2$ as a valid measure. Also, the fit is from a physical model within observation bounds, not a simple polynomial. The model residual RMS number count is 244 - the size of the dots in Figure 2. Thus, $R^2$ goodness of fit is a valid statistic here for the space expansion model.

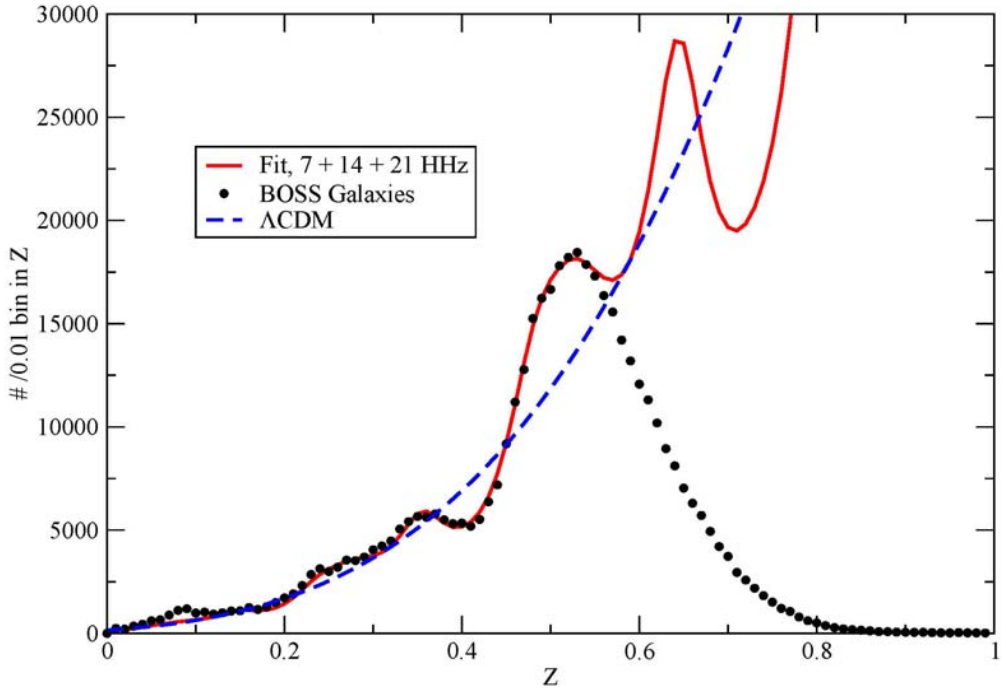

Fig.2. Space expansion model including time-shifted fundamental at ~7 HHz with 2$^{nd}$ and 3$^{rd}$ harmonics added. Dashed curve is our model with all oscillation amplitudes set to zero – effectively the ΛCDM model limit.

As was described earlier, the count dips after $z = 0.56$ and then continues to rise and peak at $z = 0.64$ while the data falls off due to target dimming. This is a prediction of the model. In fact the peaks at redshifts 0.56 and 0.64 are a periodic repeat of the two peaks between redshifts 0.2 and 0.4. We fail to include the small peak near redshift 0.1 in our fit. This may be due to the nearer structure being dominated by gravitational effects rather than the Hubble flow and could possibly be the so-called "Great Wall" which is located around redshift 0.1.

### 2.2.4 Cumulative galaxy number count

In order to get a better perspective on the distribution of galaxy number count with redshift, we shall examine the cumulative number count with increasing redshift as opposed to the binned count of Figure 2. This is done by sorting the SDSS galaxy number count with redshift increasing from zero. The result is plotted in Figure 3 as the dotted curve. The cumulative number count turns over at around 450,000 simply because the galaxies become too dim. The first thing to notice is that the count is quite smooth but has a significant "kink" at $z = 0.45$. The derivative of this curve (Figure 2) is sensitive to further, less prominent, anomalies between $z = 0.2$ and $z = 0.3$. It is stated in the DR9 release paper [8] that "In practice, it is somewhat difficult to select objects at $z = 0.45$ as the 4000Å break falls between the g and r bands. The space density of the sample at that redshift is consequently 25% lower". This would appear to be a selection problem and could account for the observed drop in number count. We have carefully examined this spectroscopic issue. The 4000Å Balmer break should redshift to 5800 Å at $z = 0.45$. This falls far enough into the r band, according to SDSS filter criteria [12], to suffer no more than 3% reduction in transmission relative to the 95% peak-average red band. So it is difficult to understand how this can account for a 25% reduction in number count. Furthermore our oscillation model accounts not only for the precisely 25%, $z = 0.45$, dip but also for all anomalies above $z = 0.1$ to high accuracy. This, however, says nothing about the source of oscillation. It could still be the result of including all the physics of dark matter and dark energy into a ΛCDM model (a damped classical oscillation in structure formation?) or it could be an oscillatory scalar field outside of ΛCDM as described in our earlier paper [2]. The fact that this LSS anomaly fundamental frequency matches the 7 HHz signal observed in earlier SNe data [2] with very different selection criteria (we showed the redshifts were completely random over their observed range) suggests there is no selection bias, but still supports either of the above oscillation-source conclusions since SNe reside in galaxies.

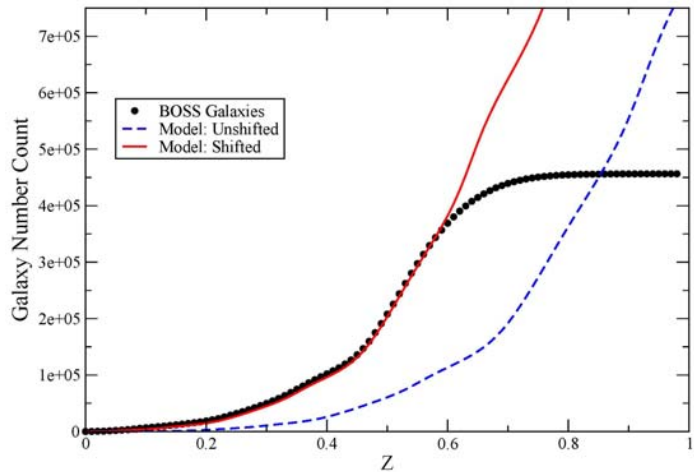

Fig.3. Cumulative galaxy number count for SDSS data comparing unshifted and time-shifted model.

## 2.3 Fourier Analysis of the SDSS Galaxy Number Count Data

If the oscillating scale factor affects the LSS, as appears to be the case in the fit of Figure 2, then a direct Fourier analysis of the SDSS galaxy number count data should reveal the same frequency spectrum. The SDSS redshift data must first be converted to lookback time from eqn. (13) and then binned in equal-time bins for a valid Fourier analysis. The model of Figure 2, based on eq. (10) but with the oscillation amplitude set to $\varepsilon = 0$, is then subtracted from the number count data - only up to $z = 0.56$ - leaving only residuals. Figure 4 shows the residuals from this model subtraction. One clear oscillation is seen between lookback times 0.3 and 0.4. This is to be compared to two full scale factor oscillations observed in [2] between lookback times 0.2 and 0.6 using SNe Ia data which were good to redshift 1.0. A single oscillation will result in a broadened frequency spectrum. A discrete Fourier analysis of Figure 2 is shown in Figure 5. The spectrum shows a strong peak at 7HHz, but is broadened as expected.

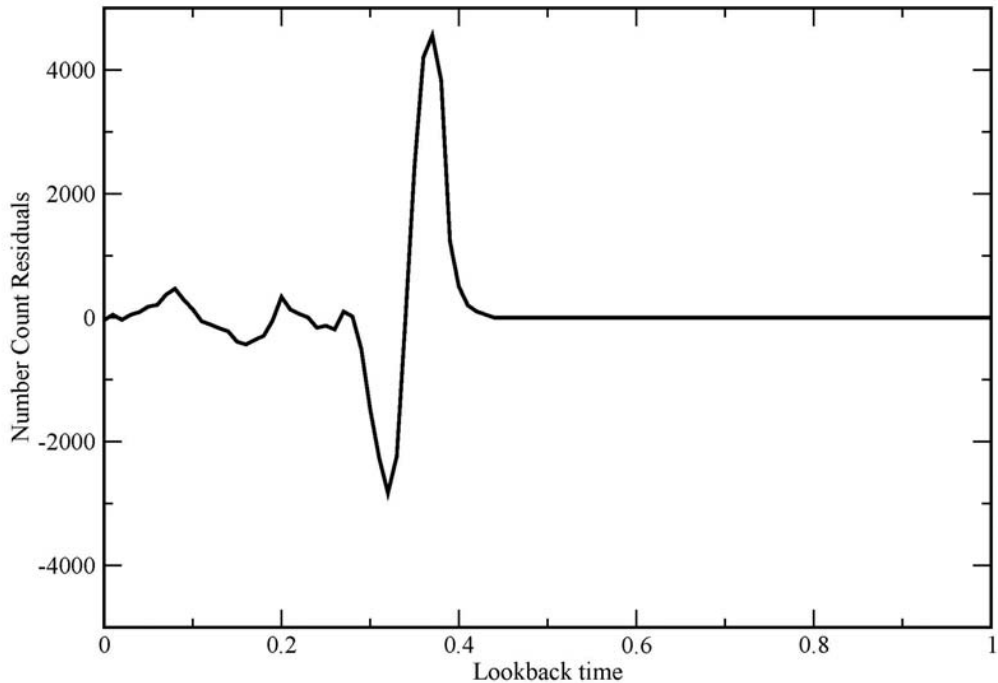

Fig.4. Number count residual oscillation in SDSS galaxy number count resulting from subtraction of our smoothed expansion number-count model from the SDSS number-count data of Fig.2. Redshift has been converted to lookback time. Lookback time 0 is the present.

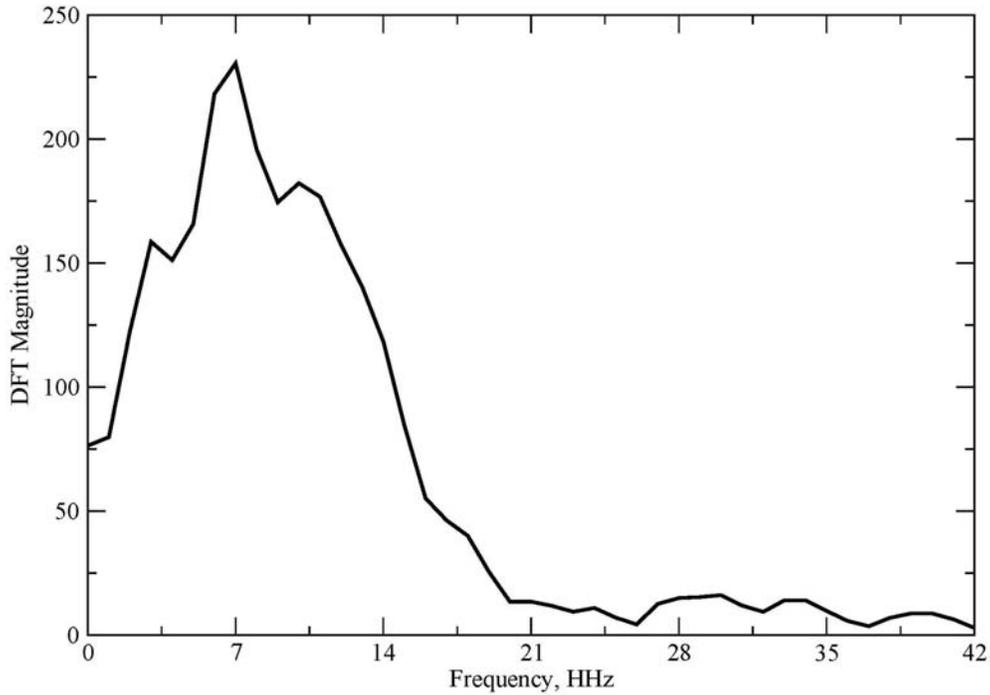

Fig.5. Discrete Fourier analysis of the signal of Figure 2. Frequency is in units of Hubble-Hertz (HHz).

## 3. CONCLUSIONS

In [2] we claim to find oscillations in the scale factor at frequencies 6.5, 13 and 20 HHz from SNe data and present a theoretical model ascribing the oscillations to scalar-field dark matter. We make a prediction that there should be an observable effect of these oscillations on the LSS as well. In the present work we incorporate the earlier results into a simple model of space expansion to assemble a number distribution of galaxies creating the observed LSS deviations described in the data release, SDSSIII-BOSS, DR9. SDSSIII-BOSS, DR10, has essentially an identical plot for twice the number count. The model using scale factor oscillations with a fundamental at ~7 HHz along with $2^{nd}$ and $3^{rd}$ harmonics fits the data at a 99.8% R-squared goodness level, once a time shift is added. The time shift can be ascribed to a delay between changes of scale factor and the onset of clustering. The scale factor governs the acceleration and deceleration of the universe. Acceleration discourages LSS formation while deceleration encourages LSS. The time shift is 0.12 Hubble-time, or approximately 1.6 Gyr. This delay is consistent with what is known about LSS formation. Furthermore, since our model is good to redshift 1, the peak at redshift 0.64 becomes a prediction. Data from a future SDSS to redshift 1 should reveal the peak before a down-turn in number due to galaxy dimming. A Fourier analysis of the unprocessed number count residuals as a function of time also shows a broadened signal sharply peaked at ~7 HHz – further reinforcing the observation of oscillations in the scale factor. Only one oscillation is observed in the SDSS data since the survey is reliable to redshift 0.56 compared to redshift 1.0 for the SNe data. Thus only the fundamental at ~7HHz is obtained – harmonics being obscured due to the broadening. Although we have shown that selection criteria are not likely to account for the observations, we cannot discount the possibility that a full scale computer simulation based on ΛCDM might explain them. That is beyond the scope of this paper.

## ACKNOWLEDGEMENTS

Funding for SDSS-III has been provided by the Alfred P. Sloan Foundation, the Participating Institutions, the National Science Foundation, and the U.S. Department of Energy Office of Science. The SDSS-III web site is http://www.sdss3.org/.

SDSS-III is managed by the Astrophysical Research Consortium for the Participating Institutions of the SDSS-III Collaboration including the University of Arizona, the Brazilian Participation Group, Brookhaven National Laboratory, Carnegie Mellon University, University of Florida, the French Participation Group, the German Participation Group, Harvard University, the Instituto de Astrofisica de Canarias, the Michigan State/Notre Dame/JINA Participation Group, Johns Hopkins University, Lawrence Berkeley National Laboratory, Max Planck Institute for Astrophysics, Max Planck Institute for Extraterrestrial Physics, New Mexico State University, New York University, Ohio State University, Pennsylvania State University, University of Portsmouth, Princeton University, the Spanish Participation Group, University of Tokyo, University of Utah, Vanderbilt University, University of Virginia, University of Washington, and Yale University.